\documentclass[aps,prl,showpacs,amsmath,amssymb,twocolumn]{revtex4-1}
\usepackage{epsfig}
\usepackage{graphicx,color}
\usepackage{amsmath}
\newcommand{\beq}{\begin{eqnarray}}
\newcommand{\eeq}{\end{eqnarray}}

\begin{document}

\title{Inclusive breakup of three-fragment weakly bound  nuclei}
\author{B. V. Carlson}
\affiliation{ Instituto Tecnol\'{o}gico de Aeron\'{a}utica, DCTA,12.228-900 S\~{a}o Jos\'{e} dos Campos, SP, Brazil }
\author{ T. Frederico}
\affiliation{ Instituto Tecnol\'{o}gico de Aeron\'{a}utica, DCTA,12.228-900 S\~{a}o Jos\'{e} dos Campos, SP, Brazil}
\author{M. S. Hussein}
\affiliation{Instituto Tecnol\'{o}gico de Aeron\'{a}utica, DCTA,12.228-900 S\~{a}o Jos\'{e} dos Campos, SP, Brazil}
\affiliation{Instituto de Estudos Avan\c{c}ados, Universidade de S\~{a}o Paulo C. P.
72012, 05508-970 S\~{a}o Paulo-SP, Brazil}
\affiliation{Instituto de F\'{\i}sica,
Universidade de S\~{a}o Paulo, C. P. 66318, 05314-970 S\~{a}o Paulo,-SP,
Brazil}


\begin{abstract}

The inclusive breakup of three-fragment projectiles is discussed within a four-body  spectator model. Both the elastic breakup and the non-elastic breakup are obtained in a unified
framework.  Originally developed in the 80's for two-fragment projectiles such as the deuteron, in this paper the theory is successfully generalized to three-fragment projectiles. The expression obtained for the inclusive cross section allows the extraction of the incomplete fusion cross section, and accordingly generalizes the surrogate method to cases such as (t,p) and (t,n) reactions. It is found that two-fragment correlations inside the projectile affect in a conspicuous way the elastic breakup cross section. The inclusive non-elastic breakup cross section is calculated and is found to contain the contribution of a three-body absorption term that is also strongly influenced by the two-fragment correlations. This latter cross section contains the so-called incomplete fusion where more than one compound nuclei are formed. Our theory describes both stable weakly bound three-fragment projectiles and unstable ones such as the Borromean nuclei.

\end{abstract}

\maketitle

Breakup reactions are of fundamental importance in collisions involving weakly bound quantum systems. \cite{BH2015}. Diatomic molecules, and two-fragment nuclei when scatter from a target may undergo a fragmentation process where one of the ions or the fragment is observed while the other fragment and the target are not. This inclusive breakup process is important as the singles spectra can supply important information about the unobserved  two-body subsystem. Here we develop a model to treat inclusive non-elastic break up reactions involving weakly bound 
three-cluster nuclei. Borromean, two-nucleon, halo nuclei are examples of unstable three-fragments 
projectiles. The model is based on the theory of inclusive breakup reactions commonly employed in 
the treatment of incomplete fusion and surrogate method. The theory was developed in the 80's by 
Ichimura, Austern and Vincent (IAV) 
\cite{Ichimura1982, IAV1985},  
Udagawa and Tamura (UT)
 \cite{UT1981} and Hussein and McVoy (HM)
\cite{HM1985}. We extend these three-body theories to derive an expression for the fragment 
yield in the reaction $A\,(a,b)\,X$, where the projectile is $a = x_1 + x_2 + b$. 
The inclusive breakup cross section is found to be the sum of a generalized four-body form of the 
elastic breakup cross section plus the inclusive non-elastic breakup cross section that involves 
the "reaction" cross section, of the participant fragments, $x_1$ and $x_2$. This latter one contains the incomplete fusion of the three-fragment projectile.\\

The final result is 
similar to the three-body case reviewed in Austern, et al. \cite{Austern1987}, 
but with important genuine four-body effects added, both in the elastic breakup cross section, which 
now contains the full correlations between the participant fragments as they scatter in the final state,  and in the inclusive non-elastic breakup, where we predict that more than one distinct compound nuclei can be formed ($x_1 + A$, $x_2 + A$, and $x_1 + x_2 + A$) in contrast to the two-fragment projectile case where only one compound nucleus is formed ($x + A$).
 Our theory of the inclusive non-elastic breakup cross section is, however,  expressed in terms of the total reaction cross sections of the three subsystems mentioned above and accordingly contain both a direct piece and capture piece. The direct piece is an inclusive cross section of all possible processes less the capture or compound nucleus part. Among the direct processes is the inelastic excitation of a state in the target, $x_1 + x_2 + A^{\star}$.These contributions can be calculated with the four-body CDCC and its totality must be subtracted in order to get the capture cross sections.\\

These developments should encourage experimentalists to seek more information about 
the $x_{1} + x_{2}$ system in the elastic breakup cross section, and the compound nuclei formed in the incomplete fusion of $a$, and for theorists to further develop and 
extend the surrogate method, based on the inclusive non-elastic breakup part of the $b$ spectrum, keeping in mind the need to calculate as precisely as possible the direct processes of the $x_1 + x_2 + A$ three-body subsystem.\\

The many-body Hamiltonian that governs the scattering dynamics of the $b + x_1 + x_2 + A$ system is 
\begin{multline}
H_{(b, x_1, x_2, A)} = [T_b + T_{{x}_1} + T_{{x}_2} + V_{b, {{x}_1}} ++ V_{b, {{x}_2}} + V_{{x}_1, {x}_2}] \\ 
+  h_A + T_A +V_{b,A} +V_{{{x}_1},A} + V_{{{x}_2},A}  
\end{multline}

where the $T$'s are the kinetic energy operators of the center of masses of the three fragments and of the target, and $h_A$ is the intrinsic Hamiltonian of the target nucleus.
We use the spectator approximation of replacing the interaction $V_{b, A}$ by the one-body optical potential $U_b$, and the no-recoil approximation $T_A$ = 0. 
The exact solution of the A + 3 many body Schr\"{o}dinger equation within the spectator model is denoted by $\varXi(\textbf{r}_b, \textbf{r}_{{x_1}}, \textbf{r}_{{x_2}}, A)$. The final channel wave function is $\chi_{b}^{(-)}(\textbf{k}_{b}, \textbf{r}_{b})\Psi^{c}_{(x_{1}x_{2}A)}$, where the distorted wave, $\chi_{b}^{(-)}$, is a solution of the optical scattering equation of the observed fragment in the presence of the optical potential $U_b$, and $c$ refers to the bound and continuum states in the $x_1 + x_2 + A$ systems. The inclusive cross section will be an integral over the coordinate of the detected, spectator fragment, $\textbf{r}_b$, and a sum over the 2 + A many-body system, $x_1 + x_2 + A$ bound and scattering states.

Using the
definition of the inclusive breakup cross section exemplified by the singles, b-spectrum and angular distribution, we can write  its exact form in the spectator model,
\begin{multline}
\frac{d^{2}\sigma_b}{dE_{b}d\Omega_{b}} = \frac{2\pi}{\hbar v}\rho_{b}(E_b)\sum_{c}\\
\cdot \left|\left\langle\chi_{b}^{(-)}\Psi^{c}_{x_{1}x_{2}A}\left|(V_{b,x_{1}} + V_{b, x_2} + V_{x_1, x_2})\right|\varXi\right \rangle\right|^{2}\\
\cdot \delta(E - E_p -E^{c})
\end{multline}
where $\rho_{b}(E_b)$ is the density of continuum states of the detected fragment, $b$, given by $\rho_{b}(E_b) \equiv [d\textbf{k}_{b}/(2\pi)^3]/[dE_{b}d\Omega_{b}]  = \mu_{b}k_{b}/[(2\pi)^{3}\hbar^3]$, where $\mu_{b}$ is the reduced mass of the $b + A$ system.\\

The connection with the four-body scattering problem hinges on developing a way where the internal coordinates of $A$ are traced out. This is accomplished by: 1) use the product approximation, $\varXi = \Psi_0^{4B(+)}\Phi_{A}$, where $\Psi_{0}^{4B (+)}$ is the exact four-body scattering wave function in the incident channel, and $\Phi_{A}$ is the ground state wave function of the target nucleus, 2) replace the delta function by the imaginary part of a Green's function which is then transformed into an operator, 3) use closure to perform the sum over $c$. We can then employ  general nuclear reaction theory accompanied by operator manipulations which exactly transform the microscopic interactions $V_{x_1, A}$ and $V_{x_2, A}$ into complex optical potentials $U_{x_1}$ and $U_{x_2}$, as done in \cite{IAV1985, HM1985, UT1981, Austern1987} to reduce the cross section into a sum of two distinct terms, the elastic breakup and the non-elastic breakup cross sections.\\
Accordingly, the inclusive breakup cross section becomes,
\begin{equation}
\frac{d^{2}\sigma_b}{dE_{b}d\Omega_{b}} = \frac{d^{2}\sigma^{EB}_b}{dE_{b}d\Omega_{b}} + \frac{d^{2}\sigma^{INEB}_b}{dE_{b}d\Omega_{b}} 
\end{equation}
where the four-body elastic breakup cross section is,
\begin{multline}
\frac{d^{2}\sigma^{EB}_b}{dE_{b}d\Omega_{b}} =  \frac{2\pi}{\hbar v_{a}}\rho_{b}(E_b)\int \frac{d{k}_{{x}_1}}{(2\pi)^{3}} \frac{d{k}_{{x}_2}}{(2\pi)^{3}}
\\
\times |\langle \chi^{3B(-)}_{{{x}_1}, {{x}_2}}\chi^{(-)}_{b}|[V_{{bx_{1}}} + V_{{bx_{2}}}] |\Psi_{0}^{4B(+)}\rangle|^2 
\\
\times  \delta(E - E_{b} - E_{({\textbf{k}_{{x}_1}}, \textbf{k}_{{x}_2})}) \label{4BEB}
\end{multline}
where $\chi^{3B(-)}_{{{x}_1}, {{x}_2}}$ is the full scattering wave function of the two unobserved fragments in the final channel. It contains the optical potentials, $U_{x_1}$, $U_{x_2}$ and the fragment-fragment interaction $V_{x_1, x_2}$ to all orders.
\begin{multline}
\frac{d^{2}\sigma^{INEB}_b}{dE_{b}d\Omega_{b}} = 
\\
\frac{2}{\hbar v_a}\rho_{b}(E_b) \langle \hat{\rho}_{{{x}_1}, {{x}_2}}|(W_{x_1} + W_{x_2} + W_{3B})|\hat{\rho}_{{{x}_1}, {{x}_2}}\rangle \label{CFH}
\end{multline}
with the source function 
\begin{multline}
\hat{\rho}_{X}(\textbf{r}_{x_{1}},\textbf{r}_{x_{2}}) = (\chi_{b}^{(-)}|\Psi_{0}^{4B(+)}\rangle = 
\\
\int d\textbf{r}_{b}\left[\chi_{b}^{(-)}(\textbf{r}_{b})\right]^{\dagger}\Psi_{0}^{4B(+)}(\textbf{r}_{b}, \textbf{r}_{x_{1}}, \textbf{r}_{x_{2}})\label{rho-4B}
\end{multline}
depending only on the coordinates of $x_1$ and $x_2$.
Once again it is important to mention that the final results contain only the optical potentials of $b$, $x_1$, and $x_2$, in so far as the interaction with the target is concerned. In Eq. ~(\ref{CFH})$, W_{x_1}$, and $W_{x_2}$ are the imaginary parts of the optical potentials of fragment $x_1$, $U_{x_1}$ and of fragment $x_2$, $U_{x_2}$, respectively.\\
Eq.(\ref{CFH}), is the four-body inclusive non-elastic breakup cross section.  We call Eq. (\ref{CFH}) the Carlson-Frederico-Hussein (CFH) formula It 
differs significantly from the three-body Austern formula \cite{Austern1987}. 
The major new features present can be quantified by writing the CFH
formula as a sum of three terms,
\begin{equation}
\frac{d^{2}\sigma^{INEB}_b}{dE_{b}d\Omega_{b}} = \rho_{b}(E_b)\sigma_{R}^{4B}
\end{equation}
\begin{equation}
\sigma_{R}^{4B} = \frac{k_a}{E_a}\left[\frac{{E_{x_1}}}{{k_{x_1}}}\sigma_{R}^{x_1} + \frac{{E_{x_2}}}{{k_{x_2}}}\sigma_{R}^{x_2} + \frac{E_{CM}({{x}_1},{{ x}_2})}{(k_{{x}_1}+ k_{{x}_2})} \sigma_{R}^{3B}\right]\label{CFH-R}
\end{equation}
where, using the form of the reaction or fusion cross section as derived in \cite{Hussein1984},
\begin{equation}
\sigma_{R}^{x_1} = \frac{{k_{x_1}}}{E_{{x_1}}} \langle \hat{\rho}_{{x}_1, {x}_2}|W_{x_1}|\hat{\rho}_{{x}_1, {x}_2}\rangle, \label{sigma_1}
\end{equation}
\begin{equation}
\sigma_{R}^{x_2} = \frac{{k_{x_2}}}{E_{{x_2}}} \langle \hat{\rho}_{{x}_1, {x}_2}|W_{x_2}|\hat{\rho}_{{x}_1, {x}_2}\rangle, \label{sigma_2}
\end{equation}
and,
\begin{equation}
\sigma_{R}^{3B} = \frac{(k_{{x}_1}+ k_{{x}_2})}{E_{CM}({{x}_1},{{ x}_2})}\ \langle \hat{\rho}_{{x}_1, {x}_2}|W_{3B}|\hat{\rho}_{{x}_1, {x}_2}\rangle \label{sigma_3B}
\end{equation}
where the energies of the different fragments are defined through the beam energy, since the projectile we are considering are weakly bound and thus the binding energy is marginally important in deciding the energies of the three fragments. Thus, e.g., $E_{{x_1},Lab} = E_{a, Lab}(M_{{x_{1}}}/M_a)$, where by $M_{a}$ and $M_{{x_1}}$ we mean the mass numbers of the projectile and fragment $x_1$, respectively.

We point out that the three-body cross section $\sigma_{R}^{3B}$ will also contain the discrete contributions of $x_1 + x_2$ bound states, if these exist. In this case, we would substitute the momentum sum defining the 3-body cross section as $k_{x_1}+k_{x_2} \rightarrow k_{CM}(x_1,x_2)$.
The formal description of an $x_1 + x_2$ bound state reaction (capture or inelastic scattering) in $\sigma_{R}^{3B}$ of the CFH theory reduces to the expression given by IAV, in which the full complexity of the three-body system is hidden in the optical potential. The 4-body formalism developed in this paper can thus take into account both the $n + p$ and the $d$ nonelastic contributions to the $\sigma_{R}^{3B}$ cross section of a $(t,n)$ reaction, for example.\\
 
The cross section, $\sigma_{R}^{x_1}$ represents the absorption of fragment $x_1$ by the target, while fragment
$x_2$ just scatters off the target through the optical potential $U_{{x}_1 A}$. The second cross section,  $\sigma_{R}^{x_2}$ is just the exchange of the role of these 
two fragments; fragment $x_2$ is captured by the target and fragment $x_1$ is scattered. Note that these cross sections are different from the one which appears in the three-body theory. The three-body sub-system $x_{1} -x_{2} -A$ is not treated within the spectator model, while the $b - x -A$  system is. Thus we anticipate that, say,  $\sigma_{R}^{{x}_1}$ in a (t,p)
reaction, will be different from $\sigma_{R}^{x}$ extracted from a (d,p) reaction.\\
It is important to  mention that the cross sections $\sigma_{R}^{x_1}$, $\sigma_{R}^{x_2}$, and $\sigma_{R}^{3B}$, would at low energies, correspond to the formation of compound nuclei of the $A +x_1$ system, the $A+x_2$ system and the the $A + (x_1 +x_2)$ system.To cite an example we take the system $^9$Be + $^{208}$Pb, whose elastic scattering  elastic breakup, and total fusion were studied recently in \cite{DDCH2015}, using a four-body Continuum Discretized Coupled Channels (CDCC) model. The $^9$Be projectile was described as a bound three fragment, $\alpha + \alpha$ + n, nucleus. Unfortunately the CDCC can not calculate the partial or incomplete fusion component of the total fusion \cite{DTT2002}. Within our four-body theory, the incomplete fusion is contained in the INEB cross section, the topic of this paper. The compound nuclei that may form in an inclusive cross section where one of the $\alpha's$ is detected, are $\alpha$ + $^{208}$Pb = $^{210}$Po, n + $^{208}$Pb = $^{209}$Pb, and $\alpha$ + n + $^{208}$Pb = $^{211}$Po, all at several excitation energies depending on where in the spectrum of the observed $\alpha$ the analysis is performed. It would be interesting to investigate experimentally the properties of these compound nuclei as they are formed in such a hybrid reaction. \\

At higher energies, these reaction cross sections may contain significant contributions from processes other than capture, such as the inelastic excitation of the target. These "direct" processes suggest writing for a given reaction cross section, the following,

\begin{equation}
\sigma_{R}^{x} = \sigma_{D}^{x} + \sigma_{CN}^{x} \label{DCN}
\end{equation}
where $x$ refers to $x_1+ A$, $x_2 + A$ or $3B = x_1+ x_2 + A$, and  $\sigma_{D}^{x}$ is the sum of inelastic cross sections involving the excitation of the target, so that the final state is $b + x_1 + x_2 +A^{\star}$. Since $b$ is the only fragment which is observed (inclusive breakup), the inelastic scattering in the three-body subsystem $x_1 + x_2 + A^{\star}$ is completely summed over, such that to an excellent approximation, $\sigma_{D}^{x} = \sum_{i} \sigma_{x + A^{\star}_{i}}$. In Eq. (\ref{DCN}), $\sigma_{CN}^{x}$ is the capture or compound nucleus (CN) cross section of the $x + A$ system.

Clearly, $\sigma_{D}^{x}$ must be theoretically well accounted for the $x_1 + A$, $x_2 + A$, and $(x_1 + x_2) + A$ subsystems with appropriate coupled channels calculation, and then subtracted from $\sigma_{R}^{x_1}$, $\sigma_{R}^{x_2}$, and $\sigma_{R}^{3B}$ to obtain the genuine capture or compound nucleus formation cross sections. \\

Other cases of particular interest are the 2n and 2p Borromean nuclei. Examples of the former are $^6$He and $^{11}$Li, while of the latter are $^{17}$Ne,  $^{20}$Mg, both having unbound (resonant) cores, $^{15}$O, and $^{18}$Ne. Inclusive $\alpha$ spectra in the breakup of, e.g.,  $^6$He would involve the formation of n +A and 2n + A compound nuclei.
It is, however, experimentally difficult to distinguish CN decay containing the same number of protons, just as in the (t, p) reaction. On the other hand the inclusive proton spectra in the breakup of,say,  $^{20}$Mg as it collides with $^{208}$Pb would involve three distinct compound nuclei; $^{209}$Bi, $^{226}$U and  $^{227}$Np, again at different excitation energies. Such an experiment would be quite challenging owing to restrictions imposed by the very short lifetimes involved and the low intensities of the secondary beam, $^{20}$Mg. Further, the fusion of unbound, resonant cores with a target, requires an investigation on its own, \cite{Dasgupta2016}. \\
The cross sections, $\sigma_{R}^{x_1}$, $\sigma_{R}^{x_2}$, of Eq. (\ref{CFH-R}) are related to the IAV cross section of the three-body theory through a convolution of the latter with the distorted wave densities, namely, $|\chi^{(+)}_{{x}_2}(\textbf{r}_{x_2})|^2$, 
and $|\chi^{(+)}_{{x}_1}(\textbf{r}_{x_1})|^2$, respectively. This can be easily seen if an eikonal-type approximation of the projectile distorted wave is used. The incident wave function, in the naive (see discussion below) product approximation of the four-body wf, in the DWBA, is the product of the distorted wave of the projectile, $\chi_{a}^{(+)}$ times the intrinsic projectile wf, $\Phi_a({r_b},{{{r_{x_{1}}}}}, {{{r_{x_{2}}}}})$. Thus all matrix elements involving the incident channels wf will be constrained by $\Phi_a({{r_b}},{{{r_{x_{1}}}}}, {{{r_{x_{2}}}}})$.

The above distorted wave densities,  $|\chi^{(+)}_{{x}_2}(\textbf{r}_{x_2})|^2$, 
and $|\chi^{(+)}_{{x}_1}(\textbf{r}_{x_1})|^2$, arise from the solutions of a non-Hermitian Schr\"{o}dinger equation with the respective optical potentials. In Ref. \cite{Hussein1987}, these distorted wave densities were calculated and found to be related to the reaction cross section of $x_1$ and of $x_2$. To be more specific, we first write the source function $\hat{\rho}^{4B}_{HM}$, Eq. (\ref{rho-4B})
\begin{multline}
\langle \textbf{r}_{{x}_1}, \textbf{r}_{{x}_2}|\hat{\rho}^{4B}_{HM}\rangle = 
\\\int d\textbf{r}_{b} \Phi_{a}(\textbf{r}_{{x}_1}, \textbf{r}_{{x}_2}, \textbf{r}_b) \langle\chi^{(-)}_{b}|\chi^{(+)}_{b}\rangle(\textbf{r}_b) 
\\
\times \chi^{(+)}_{{x}_1}(\textbf{r}_{{x}_1})\chi^{(+)}_{{x}_2}(\textbf{r}_{{x}_2})
\end{multline}
The overlap function $\langle\chi^{(-)}_{b}|\chi^{(+)}_{b}\rangle(\textbf{r}_b) $ is the integrand of the elastic S-matrix element of the spectator fragment, $b$, $S_{\textbf{k}_{b}^{\prime}, \textbf{k}_{b}}=\int d\textbf{r}_{b} \langle\chi^{(-)}_{b}|\chi^{(+)}_{b}\rangle(\textbf{r}_b)$.  Thus, we introduce the internal motion modified S-matrix of the b fragment, $\hat{S}_{b}(\textbf{r}_{{x}_1}, \textbf{r}_{{x}_2}) \equiv \int d\textbf{r}_{b} \Phi_{a}(\textbf{r}_{{x}_1}, \textbf{r}_{{x}_2}, \textbf{r}_b)\langle\chi^{(-)}_{b}|\chi^{(+)}_{b}\rangle(\textbf{r}_b) $. Accordingly, the source function becomes,
\begin{equation}
\langle \textbf{r}_{{x}_1}, \textbf{r}_{{x}_2}|\hat{\rho}^{4B}_{HM}\rangle = \hat{S}_{b}(\textbf{r}_{{x}_1}, \textbf{r}_{{x}_2})\chi^{(+)}_{{x}_1}(\textbf{r}_{{x}_1})\chi^{(+)}_{{x}_2}(\textbf{r}_{{x}_2})
\end{equation}
The cross section, $\sigma_{R}^{x_1}$, Eq. (\ref{sigma_1}), becomes,
\begin{multline}
\frac{E_{{x}_1}}{k_{{x}_1}}\sigma_{R}^{x_1} = \int d\textbf{r}_{{x}_1}d\textbf{r}_{{x}_2} |\hat{S}_{b}(\textbf{r}_{{x}_1}, \textbf{r}_{{x}_2})|^{2}|\chi^{(+)}_{{x}_2}(\textbf{r}_{x_2})|^2
\\
\times W(\textbf{r}_{{x}_1})|\chi^{(+)}_{{x}_1}(\textbf{r}_{x_1})|^2 \label{psi4bapp-13}
\end{multline}
Similar expression is found for the cross section, $\sigma_{R}^{x_2}$, Eq. (\ref{sigma_2}), namely,

\begin{multline}
\frac{E_{{x}_2}}{k_{{x}_2}}\sigma_{R}^{x_2} = \int d\textbf{r}_{{x}_1}d\textbf{r}_{{x}_2} |\hat{S}_{b}(\textbf{r}_{{x}_1}, \textbf{r}_{{x}_2})|^{2}|\chi^{(+)}_{{x}_1}(\textbf{r}_{x_1})|^2
\\
\times W(\textbf{r}_{{x}_2})|\chi^{(+)}_{{x}_2}(\textbf{r}_{x_2})|^2 \label{psi4bapp2-13}
\end{multline}

to be compared to the two-body reaction cross section of the fragment $x$, in the breakup of a two-cluster projectile, $a = x + b$,
\begin{equation}
\frac{E_{x}}{k_x}\sigma_{R}^{x} = \int d\textbf{r}_{x} |\hat{S}_{b}(\textbf{r}_x)|^{2} W(\textbf{r}_x)|\chi^{(+)}_{x}(\textbf{r}_{x})|^2
\end{equation}
where $ \hat{S}_{b}(\textbf{r}_x) \equiv \int d\textbf{r}_{b}  \langle\chi^{(-)}_{b}|\chi^{(+)}_{b}\rangle(\textbf{r}_b)\Phi_{a}(\textbf{r}_b, \textbf{r}_x)$. 
One sees clearly that the 4B cross sections, $\sigma_{R}^{x_1}$ and $\sigma_{R}^{x_2}$ are damped compared to the 3B one owing, among other factors, to the presence of the distorted 
wave  densities $|\chi^{(+)}_{{x}_2}(\textbf{r}_{x_2})|^2$, and $|\chi^{(+)}_{{x}_1}(\textbf{r}_{x_1})|^2$ in the former ones. It is important to remind once again that the internal intrinsic wave function of the projectile is present in these formulae through the modified $b$- S-matrix factors, $|\hat{S}_{b}(\textbf{r}_{{x}_1}, \textbf{r}_{{x}_2})|^2$, and $|\hat{S}_{b}(\textbf{r}_x)|^{2}$.
Finally It is also instructive to compare the above cross sections to the "free" one, where x is the primary projectile,
\begin{equation}
\frac{E}{k}\sigma_{R} = \int \textbf{dr} |\chi^{(+)}(\textbf{r})|^2 W(\textbf{r})
\end{equation}

\begin{figure}[htb]
  \centerline{\epsfig{figure=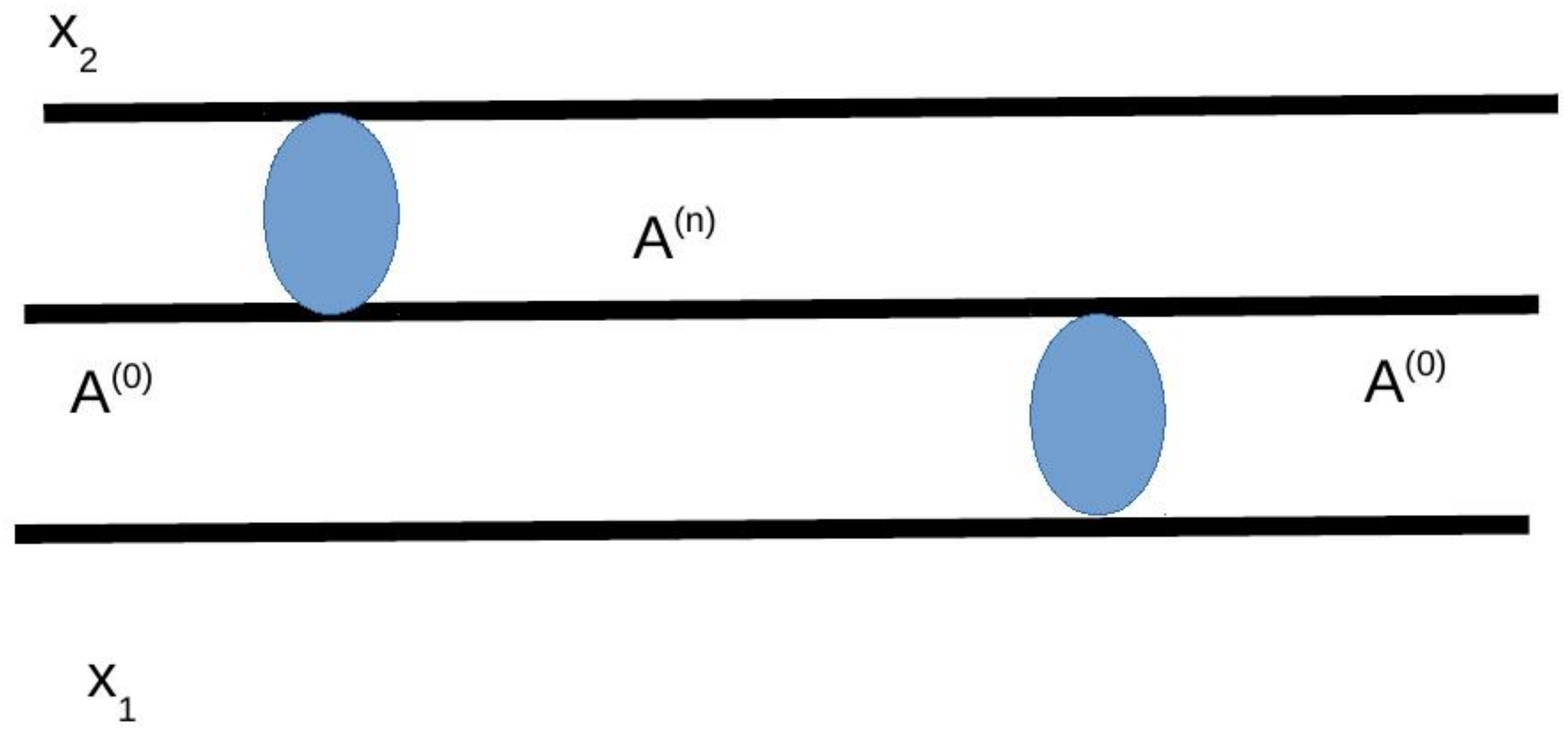,height=6cm}}
\vspace{-1cm}
\caption{Three-body optical potential $U_{3B}$. Excitation of the target by particle $x_1$ and 
de-excitation by $x_2$ (see Eq. (\ref{U3B})).}
\label{fig1}
\end{figure}

Finally the last cross section, $\sigma_{R}^{3B}$, is new and a genuine three-body absorption cross section. In the following we take a critical look at its structure.
Using the wisdom of conventional nuclear reaction theory, the three-body $W_{3B}$ results from the average of processes involving the virtual excitation of the target by one of the fragment and its virtual de-excitation by the other fragment, as well as other processes, with the final result being the full capture, or complete fusion, of both fragments, as illustrated in Fig. \ref{fig1},

In the language of projection operators, the 3B optical potential, $U_{3B}$, whose imaginary part is -$W_{3B}$, is given by
\begin{multline}
U_{3B} = PV_{{x_1}A}Q (QG_{{{x}_1}{{x}_2}A}(E_{x})Q) QV_{{x_2}A}P + 
\\
PV_{{{x}_2}A}Q (QG_{{{x}_1}{{x}_2}A}(E_{{{x}_1}{{x}_2}})Q) QV_{{{x}_1}A}P \label{U3B}
\end{multline}
where,  the Q-projected 3B Green's function of the $x_1 + x_2 + A$ system, $QG_{{{x}_1}{{x}_2}A}(E_{{{x}_1}{{x}_2}})Q\equiv QG_{{{{x}_1}{{x}_2}A}}Q$, is given by
\begin{small}
\begin{multline}
QG_{{{x}_1}{{x}_2}A}Q = 
\\
\frac{1}{E_{{{x}_1}{{x}_2}} - QH_{0}Q + Q[V_{{{x}_1}A}+ V_{{{x}_2}A}]PG_{0}P[V_{{{x}_1}A} + V_{{{x}_2}A}]Q + i\varepsilon}
\end{multline}
\end{small}
The imaginary part of $U_{3B}$ is now easily calculated. \begin{multline}
Im[QG_{{{x}_1}{{x}_2}A}Q] = -\pi \Omega^{(-)}_{Q}\delta(E_x - QH_{0}Q)(\Omega^{(-)}_{Q})^{\dagger} +
\\  
+(QG_{{{x}_1}{{x}_2}A}Q)^{\dagger} 
\\
\times Q[V_{{{x}_1}A} + V_{{{x}_2}A}]\,P\delta(E_{{{x}_1}{{x}_2}}- PH_{0}P)P\,[V_{{{x}_2}A} + V_{{{x}_2}A}]Q
\\
\times QG_{{{x}_1}{{x}_2}A}Q   
\end{multline}

Thus,
\begin{multline}
W_{3B} = \pi[PV_{{x_1A}}Q \Omega^{(-)}_{Q}\delta(E_x - QH_{0}Q) 
\\
(\Omega^{(-)}_{Q})^{\dagger}QV_{{x_2A}}P  +(x_{1}\leftrightarrow x_{2})+
\\
+ PV_{{x_1A}}Q (QG_{{{x}_1}{{x}_2}A}Q)^{\dagger} Q[V_{{x_1A}} + V_{{x_2A}}]P
\\
P\delta(E_x - PH_{0}P)P[V_{{x_1A}}+V_{{x_2A}}]Q  (QG_{{{x}_1}{{x}_2}A}Q) QV_{{x_2A}}P+
\\
 + PV_{{x_2A}}Q (QG_{{{x}_1}{{x}_2}A}Q)^{\dagger} Q[V_{{x_1A}} + V_{{x_2A}}]P
 \\
 P\delta(E_x - PH_{0})P)P[V_{{x_1A}} + V_{{x_2A}}]Q  (QG_{{{x}_1}{{x}_2}A}Q)  QV_{{x_1A}}P] 
\end{multline}

Therefore the reactive content of $W_{3B}$ is simple to discern. The first term corresponds to the already announced virtual excitation of the target by one fragment followed by a virtual de-excitation through the action of the second fragment. The last two terms corresponds to absorption of the two fragments by the target. There are eight  terms which describe the different ways this absorption is manifested. It is evident that a detailed evaluation of $U_{3B}$ is a formidable task. The correlation is induced by the interaction $V_{{x_1}{x_2}}$, which besides scattering the two fragments, could bind them in a resonance or quasi-bound state. Accordingly we replace the very complicated structure above by a simple effective two-body fusion. This applies to the calculation of the reaction cross section, $\sigma^{3B}_{R}$.  Of course, we can settle on less and use the experience acquired over several decades in the description of complete fusion \cite{CGDH2006, CGDLH2015} and treat the cross section  $\sigma_{R}^{3B}$, Eq.(\ref{sigma_3B}), as the capture or complete fusion of the system $x_1 + x_2$ by the target, modified by the internal motion of the two fragments inside the projectile. For the purpose of illustration we use the eikonal-type approximation of the DWBA version of the four-body wave function. This entails using $\Psi^{(+ 4B)} \approx \chi^{(+)}_{b}(\textbf{r}_b)\Psi^{(+)}_{{{x}_1}, {{x}_2}}(\textbf{r}_{{x}_1}, \textbf{r}_{{x}_2})\Phi_{a}(\textbf{r}_{b}, \textbf{r}_{{x}_1}, \textbf{r}_{{x}_2})$. This then allows writing,
\begin{multline}
\sigma_{R}^{3B} = \sigma_{CF}^{3B} =  \frac{(k_{{x}_1}+ k_{{x}_2})}{E_{CM}({{x}_1},{{ x}_2})}
\langle \hat{\rho}_{{x}_1, {x}_2}|W_{3B}|\hat{\rho}_{{x}_1, {x}_2}\rangle
\\
= \frac{(k_{{x}_1}+ k_{{x}_2})}{E_{CM}({{x}_1},{{ x}_2})}\int d\textbf{r}_{{x}_1}d\textbf{r}_{{x}_2} |\hat{S}_{b}(\textbf{r}_{{x}_1}, \textbf{r}_{{x}_2})|^{2}
\\
\times \left|\Psi^{(+)}_{{{x}_1}, {{x}_2}}(\textbf{r}_{{x}_1}, \textbf{r}_{{x}_2})\right|^{2}W_{3B}(\textbf{r}_{{x}_1}, \textbf{r}_{{x}_2})\, ,
\end{multline}
where the correlation between the $x_1$ and $x_2$ fragments are kept in the three-body scattering wave function with the target.

In this paper we have derived the 3-fragment projectile inclusive breakup cross section and pointed out the major differences from the corresponding cross section in the case of two-fragment projectile currently used in calculations. Our theory permits the study of fragment-fragment correlations through a judicious coincidence measurement of the elastic breakup part of the cross section. For the inclusive non-elastic breakup, or incomplete fusion, part of the cross section, we have derived formulae reminiscent of the so-called Austern formula, with two major differences. Our 4-body formula contains reference to the three-body nature of the fusing two fragments and to the intrinsically three-body "direct" process, which permits the virtual excitation of the target by one of the fragments followed by the target de-excitation by the other fragment. . 
The imaginary part of the optical potential is found to be composed of the sum of two one-fragment potentials, plus a new, 3-body part, which contains the fusion of the two fragments. We propose a simplified model to deal with this three-body absorption term in the imaginary part of this latter potential. A simplified treatment of $\sigma_{R}^{3B}$, such as treating the two fragments that fuse with target as one, di-fragment, and the compound nucleus formed is of the system $A + (x_1 +x_2)$, our four-body theory still maintains its premise of being so, as it allows the formation of the compound nuclei of the systems $A+ x_1$, and $A+x_2$. In a Surrogate model of reactions of the type (t, p), the compound nuclei formed, as emphasized above, would be $A + n$, and $A+2n$. If the triton were to be treated as a two fragment projectile composed of a proton bound to a di-neutron, then wrong conclusions (over estimate) about the compound nucleus, $A + 2n$, formation strength would be reached, as the $A +n$ compound nucleus predicted by our theory will be completely missed. Our results should be quite useful in the study of inclusive breakup of unstable thee-fragments projectiles, such as the Borromean nuclei, where two neutrons or two protons are involved in the reaction mechanism. 
Hybrid theories, such as the Surrogate Method, \cite{Escher2012}, can now be extended to the case of, say, tritium breakup.\\

The DWBA version of the theory is also developed. Such a distorted wave approximate requires the employment of the four-body Faddeev-Yakubovsky equations \cite{Faddeev1961, Yakubov1967}, just as the three-body theory requires the three-body Faddeev equations \cite{Austern1987, HFM1990} for its DWBA limit. The F-Y reduction has been accomplished in reference \cite{CFH2016}. In a nut shell these equations act as a guide to obtain the correct DWBA limit of our four-body theory. It is important to remind the reader that the DWBA approximation of the full four-body scattering wave function is not just replacing it by distorted waves, but rather identify the dominant component of the 18 coupled F-Y equations (for four non-identical particles) and use the distorted wave in this component. The four-body wave function is then replaced by the DWBA approximation of this dominant F-Y wave function. This allows obtaining the IAV version of our four-body problem. The simple replacement of the four-body wave function by the distorted wave of the projectile times the ground state wave function of the $a$ supplies only the non-orthogonality or HM term which, when used in the development of the source function, gives a sum rule involving the post, IAV piece, and the UT piece,

\begin{equation}
\hat{\rho}_{x_1, x_2}^{IAV} = \hat{\rho}_{x_1, x_2}^{UT} +\hat{\rho}_{x_1, x_2}^{HM} \label{rhoT}
\end{equation}
where,
\begin{equation}
\hat{\rho}_{x_1, x_2}^{IAV} \equiv <\chi_{b}^{(-)}|G^{(+)}_{b, x_1, x_2, A}\left[V_{b,x_1} + V_{b,x_2}\right]|\chi^{(+)}_{a}\Phi_{a}> \label{rhoIAV}
\end{equation}
\begin{equation}
\hat{\rho}_{x_1, x_2}^{UT} \equiv G^{(+)}_{x_1, x_2, A}<\chi_{b}^{(-)}|\left[U_b + U_{x_1} + U_{x_2} - U_a\right]|\chi^{(+)}_{a}\Phi_{a}> \label{rhoUT}
\end{equation}
and
\begin{equation}
\hat{\rho}_{x_1, x_2}^{HM} = <\chi_{b}^{(-)}|\chi_{a}^{(+)}\Phi_{a}> \label{rhoHM}
\end{equation}
where the Green's function $G^{(+)}_{x_1, x_2, A} = [E - E_b - T_{x_1} - T_{x_2} - V_{x_1, x_2} - U_{x_1} - U_{x_2} + i\varepsilon]^{-1}$, while $G^{(+)}_{b, x_1, x_2, A} =
[E - T_{b} - T_{x_1} - T_{x_2} - V_{x_1, V_2} - U_{b} - U_{x_1} - U_{x_2} + i\varepsilon]^{-1}$. 

The above results, Eq.~(\ref{rhoT}, \ref{rhoIAV}, \ref{rhoUT}, \ref{rhoHM}), confirms that the general structure of the CFH cross section, Eq.~(\ref{CFH}), is, in the DWBA limit, similar to the three-body case, with the full post form (or the four-body IAV) can be written as the sum of the prior four-body UT cross section plus the four-body HM one plus the interference term \cite{Ichimura1990, HFM1990}. The major difference between the 4B and 3B cases resides in the structure of the reaction cross sections for the absorption of one of the interacting fragments, which we find to be damped by the absorption effect of the other fragment. In general we expect that in cases such as the (t,p) reaction, the one neutron absorption cross section will be smaller than the corresponding one in the (d,p) reaction. Another important new feature which has already been alluded to above is the presence of the three-body absorption term.

In the case of a two-fragment projectile, the $x_1 + x_2$, becomes just $x$, and the above source functions are used in the calculation of, say, the (d,p) cross section as was done by \cite{Potel2015} using the prior form, $\rho_{x}^{UT}$, and by \cite{Ducasse2015, Moro2-2015, Carlson2015}, using the post IAV form $\rho_{x}^{IAV}$. Recently, \cite{Moro2016}, extended the calculation to the case of $^6$Li (= $\alpha$ + d) breakup on several targets. In their calculation, \cite{Moro2016} used the CDCC for the elastic breakup and the IAV cross section for the INEB. The summed cross section was found to be in very good agreement with the measured $\alpha$ spectra. It would be exceedingly interesting to extend such calculations to $^9$Be inclusive $\alpha$ spectra using the four-body formalism developed in this Letter and exemplified by the CFH equation, Eq. (\ref{CFH}), for the INEB cross section.
Further, the interaction of two correlated neutrons with the target, as measured by the cross section $\sigma^{3B}_{R}$, could lead to the excitation of Giant Pairing Vibration (GPV) resonances, predicted in \cite{Broglia1977} and experimentally studied in light targets quite recently \cite{Cappuzz2015}. These collective states involve the coherent excitation of particle-particle pairs, in complete analogy to the coherent excitation of particle-hole pairs that constitutes the microscopic foundation of multipole giant resonances. The potential excitation of the GPV opens interesting prospects for nuclear structure studies in reactions of the type (t, p), the two-neutron Borromean cases, ($^{6}$He, $^4$He), ($^{11}$Li, $^9$Li), ($^{14}$Be, $^{12}$Be), ($^{22}$C, $^{20}$C), and the two-proton halo cases, ($^{17}$Ne, $^{15}$O), and
($^{20}$Mg, $^{18}$Ne). The theory we have developed in this paper would be the appropriate framework in which to study these types of collective nuclear excitations, associated with pairing correlations in the target.

{\it Acknowledgements.} 
This work was partly supported by the Brazilian agencies, Funda\c c\~ao de Amparo \`a Pesquisa do Estado de
 S\~ao Paulo (FAPESP), the 
Conselho Nacional de Desenvolvimento Cient\'ifico e Tecnol\'ogico  (CNPq). MSH also acknowledges a Senior Visiting Professorship granted by the Coordena\c c\~ao de Aperfei\c coamento de Pessoal de N\'ivel Superior (CAPES), through the CAPES/ITA-PVS program.

\end{document}